\documentclass[twocolumn, showpacs, amsmath, amssymb, prb]{revtex4}

\usepackage{amssymb}
\usepackage{amsmath}
\usepackage{amsfonts}
\usepackage{graphicx}

\makeatother

\newcommand{\be}{\begin{equation}}
\newcommand{\ee}{\end{equation}}
\newcommand{\ba}{\begin{eqnarray}}
\newcommand{\ea}{\end{eqnarray}}

\begin{document}

\title{Resonance plasmon linewidth oscillations in spheroidal metallic nanoparticle embedded in a dielectric matrix}

\author{Nicolas I.~Grigorchuk \footnote{email: ngrigor@bitp.kiev.ua}}

\affiliation{Bogolyubov Institute for Theoretical Physics, National Academy of Sciences of Ukraine, \\
 14-b Metrologichna Str., Kyiv-143, Ukraine, 03680}

\begin{abstract}

The kinetic approach is applied to calculate oscillations of a surface plasmon linewidth in a spheroidal metal nanoparticle embedded in any dielectric media. The principal attention is focused on the case, when the free electron path is much greater than the particle size.
The linewidth of the plasmon resonance as a function of the particle radius, shape, dielectric constant of the surrounding medium, and the light frequency is studied in detail. It is found that the resonance plasmon linewidth oscillates with increasing both the particle size and the dielectric constant of surrounding medium. The main attention is paid to the electron surface-scattering contribution to the plasmon decay. All calculations the plasmon resonance linewidth are illustrated by the example of the Na
nanoparticles with different radii. The results obtained in the kinetic approach are compared with the known ones from other models.
The role of the radiative damping is discussed as well.

\end{abstract}

\pacs{78.67.Bf; 68.49.Jk; 73.23.-b; 78.67.-n; 52.25.Os; 36.40.Vz; 25.20.Dc; 52.25.Os; 78.47.+p.}

\date{\today}

\maketitle


\newpage
\section{Introduction}
\label{intro}

The surface plasmon (SP) excitation (dipolar resonance) in metallic nanoparticles (MNs) are still of great fundamental interest\cite{WAH,MB,BB,BH,Hu,KV} since its pronounced local resonances, the position, shape and intensity can be tuned over wide spectral range by varying the size and shape of the MN or by changing the surrounding medium. Once excited, plasmon oscillations can
damp non-radiatively by absorption caused by electron-phonon interactions, and/or radiatively by the resonant scattering process.
Due to their interaction with other internal degrees of freedom, the plasmon excitations can decay that results in a certain
linewidth $\Gamma$ [full width at half a maximum (FWHM)].
The linewidth of the plasmon resonances are an important parameter, since it contains the information about
the character of interactions in the system, in particular, about the lifetime of the surface plasmon.
Furthermore, the information on $\Gamma$ would help to analyze the specificity of electronic decay in the MNs.

The interest to the study the plasmon linewidth is maintained because it is connected to the local field enhancement,
an effect which increases the intensity of the incident light near the MNs surface by several orders of magnitude.\cite{BH,Hu,KV}
Since a lot of devices incorporating MNs gains from this effect, $\Gamma$ is treated as a main parameter in applications
such as field concentration for nanopatterning with nanowires,\cite{LX} plasmonic nanolithography,\cite{SFS} and near-field
optical microscopy,\cite{SD} astigmatic optical tweezers,\cite{APS} surface enhanced Raman scattering,\cite{Nie} etc.

The damping of surface plasmon resonances in MNs has been well studied both
experimentally\cite{KG,BCL,PBL,KPG,LS,SBW,WMP,SFW,MBS,BHS,ACD,BCT,HNF} and theoretically\cite{KK,YBB,MWJ,YB,BS,WMW,WWI,P,CS}.
Usually, both the bulk and the surface damping mechanisms play an important role in the surface plasmon decay.
In the MNs of smaller radii, the penetration depth of the plasmon field reduces and becomes more localized near the surface.
As a result, the bulk-induced loss processes play only a minor role and
the electronic excitations generated by the surface potential dominate.

In the bulk material, the damping constant $\Gamma$ is related to the lifetimes of all electron scattering
processes that are mainly due to inelastic electron-electron, electron-phonon, an electron-defect scattering,
Landau damping, an excitation of bound electrons into the conduction band, emission of electrons,\cite{KV,AM} etc.
However, if electrons are confined in nanoparticles with sizes below 100\AA,
the electron-surface scattering comes into play.\cite{SBW}
The surface acts as an additional scatterer because the mean free path (MFP) of the electrons becomes comparable to the size of the particles. For very small particles, the collisions of conduction electrons with the particle surface dominate, that results
in the reduction of effective MFP. According to this model, the damping constant $\Gamma$ depends on the particle radius R.
The width of the spectral line has been often explained\cite{KV} by introducing a term varying linearly with the inverse
diameter of the particle.

Kreibig was the first who introduced this classical picture of the limitation of the MFP and found good agreement
with experimental results.\cite{KV} An early quantum mechanical model developed by Kawabata and Kubo\cite{KK} also predicts
a 1/R dependence of the plasmon bandwidth. However, their model do not treat the surface as a scatterer for the electrons.
The 1/R dependence reflects the importance of the ratio between the surface area $S$ and the volume $V$.

In Refs.~\onlinecite{MWJ,YB,WMW,WWI}, the semiclassical theory was used to evaluate the surface plasmon linewidth in MNs.
The oscillations of the surface plasmon lifetime as a function of nanoparticle size have been found with the use of such approaches.
By numerical calculations, it was found that the nonmonotonous size-dependent structure in the line shape can be attributed
to the shell effects.  For small alkaline-metal clusters has been shown that the angular-momentum-dependent electron-hole
density-density correlations lead to an oscillatory size-dependence of the plasmon linewidth due to Landau damping.

But not only these correlations and Landau damping cause the linewidth oscillations.
The purpose of the present paper is to calculate the linewidth of the surface-plasmon resonance when the MFP
of the electrons is vastly larger than the particle size and their scattering on the particle surface plays the most important role.
Within the framework of the kinetic approach, we found that the the surface plasmon linewidth exhibits the oscillations as a function
of the nanoparticle size, shape, and light frequency. We will demonstrate how they correspond to the oscillations detected earlier\cite{MWJ,YB,WMW,WWI} in numerical calculations based on the time dependent local density approximation.

Therefore, the specific interest is to study the behavior of the resonance linewidth with varying of such a parameters
as the particle radius, its shape, the dielectric constant of surrounded medium, the deviation of light frequency from
the frequency of plasmon resonance and some others. The dielectric constant of the MN bulk is taken into account as well.

The rest of the paper is organized as follows.
The theoretical background to the problem is presented in Section II.
Section III contains the study the plasmon linewidth in some specific cases.
Section IV is devoted to the discussion of the obtained results.
The radiative damping is studied in the Section V and
Sections VI contains the summary.

\section{Theoretical background}

The interaction of light with a MN embedded in the medium is studied in the framework of classical optics,
assuming that the particle and the medium are continuous, homogeneous, and characterized by their dielectric function.
To overcome the problem connected to the inhomogeneous line broadening (connected to the size and
shape distribution within the particle ensemble), we will restricted ourselves only to a {\em single} MN,
which directly yields the homogeneous linewidth.
The single-particle scattering spectroscopy based on the localized surface plasmon
resonance spectra of a single metal nanoparticle is well developed now.\cite{KPG,SFW,ACD,BCT}

In the classical case of free electrons in bulk metal, the damping $\gamma_b (\equiv\nu)$ is
due to the inelastic scattering of the electrons with phonons, lattice defects, or impurities
($\nu$ refers to the electron collision frequency), which shorten the MFP.
In this case, the relation $\gamma_b=\upsilon_F/l_{\infty}$ holds, where $\upsilon_F$ is the Fermi velocity
and $l_{\infty}$ is the MFP of conduction electrons in the bulk.
Similarly, to estimate the surface effect in electron-surface scattering,
the following empirical relation\cite{KV,YB,CS,BCT}
\be
 \label{eq 1}
  \gamma_s=A\frac{\upsilon_F}{l_{\rm eff}}
   \ee
is often used, where $l_{\rm eff}$ is reduced effective MFP and $A$ is a phenomenological factor.
But such a formula can be applied to the MNs of a spherical shape only in the case when the MFP
of the electrons $l$ is smaller than the particle size $d$.
If the shape of MN differs from the spherical one and $l\gg d$, then the formula (\ref{eq 1}) can no longer be used.

 We will consider the scattering of electromagnetic (EM) waves on metallic nanoparticle, the size of which $d$
 is much smaller than the wavelength of light $\lambda\sim c/\omega$, or $kd\ll 1$, where $c$ is the speed
 of light in vacuum, $\omega$ is the angular frequency of the light and $k$ refers to the wave number.
Then the EM field around the MN can be considered to be homogeneous and across a particle as uniform,
such that all the conduction electrons move in-phase producing only dipole-type oscillations.
An additional limitation to the particle size is connected to the electron MFP.
The MN size we take as being much smaller than $l$. Then collisions of the conduction
electrons with the particle surface become the most important relaxation process.
A diffuse boundary scattering is assumed to be a good approximation in this case.

Let us consider the MN embedded in a homogeneous, transparent medium with a dielectric constant $\epsilon_m$.
The average power absorbed by the MN from the external EM field in the dipolar approximation is given by\cite{TG,GT1}
\be
 \label{eq 2}
  W=\frac{V}{2}\sum_{j=1}^3 \sigma'_{jj} |E_{j,{\rm in}}|^2,
   \ee
where $V$ is the particle volume, $\sigma'_{jj}$ is the principal components of the real part of the conductivity tensor,
$E_{j,in}$ are the components of the electric field inside the MN, which are connected to the components
of the uniform external electric field (coordinate independent) $E_j^{(0)}$ by the relation \cite{LL}
\be
 \label{eq 3}
  E_{j,{\rm in}}({\bf r}) = \frac{E_j^{(0)}(0)}{1+L_j(\epsilon/\epsilon_m -1)},
   \ee
where $L_j$ are the principal value of the $j$-th component of the depolarization tensor
that is also known as a geometric factor. The explicit expressions of $L_j$ for a MN with a particular
shape can be found elesewhere (see, e.g., Refs.~[\onlinecite{Osb}], [\onlinecite{BH}],  and [\onlinecite{LL}]).
By $\epsilon(=\epsilon'+\epsilon'')$, the complex dielectric permittivity of the particle material, which would
correspond to a given frequency $\omega$, is denoted. Substituting Eq.~(\ref{eq 3}) in Eq.~(\ref{eq 2}), one finds
\be
 \label{eq 4}
  W=\frac{V}{2}\sum_{j=1}^3\frac{
   \sigma'_{jj}(\epsilon'^2_m+\epsilon''^2_m)\, |E_j^{(0)}|^2}{[\epsilon'_m+L_j(\epsilon'-\epsilon'_m)]^2+[\epsilon''_m+L_j(\epsilon''-\epsilon''_m)]^2}.
    \ee
Here $\epsilon'$ and $\epsilon''$ are, respectively, the real and imaginary parts of the dielectric permittivity of the particle material; $\epsilon'_m$ and $\epsilon''_m$ are, respectively, the real and imaginary parts of the dielectric permittivity of the matrix.

To account the surface effect, the dielectric function\cite{HNF,CS}
\be
 \label{eq 5}
  \begin{array}{ll}&\epsilon'(\omega,l_{\rm eff})=
   \epsilon_{\rm inter}(\omega)+\left(1-\frac{\omega^2_{pl}}{\omega^2+(\gamma_b+\gamma_s)^2}\right),
    \\ \\&\epsilon''(\omega,l_{\rm eff})=\frac{\omega^2_{pl}}{\omega}
     \frac{\gamma_b+\gamma_s}{\omega^2+(\gamma_b+\gamma_s)^2}\end{array}
      \ee
is used often, where $\epsilon_{\rm inter}$ accounts for the {\em inter}band electron transitions, and the expression
in the parentheses applied for the {\em intra}band electron transitions. The effect of the surface is reduced simply to the
addition of a term $\gamma_s \equiv\gamma_s(l_{\rm eff})$ in the denominators of Eq.~(\ref{eq 5}) in the form of Eq.~(\ref{eq 1}).

In our model we will seek the expression for $\gamma_{s}$ directly from the kinetic equation method.
We restrict ourselves to the case of frequency close to the frequency of the bulk plasma
oscillations of electrons in metal [$\omega_{\rm pl}=(4\pi n e^2/m)^{1/2}$].
Then both terms $\gamma_b$ and $\gamma_s$ can be neglected in the denominators of Eq.~(\ref{eq 5}) and the
dielectric function for a free electron gas can be expressed within the Drude-Sommerfeld model as\cite{BH,KV,AM,GT1}
\be
 \label{eq 6}
  \epsilon'_{jj}(\omega)= 1-\frac{\omega^2_{\rm pl}}{\omega^2}, \qquad
   \epsilon''_{jj}(\omega)=4\pi\frac{\sigma'_{jj}(\omega)}{\omega}.
    \ee
The effect of a core polarization (the {\em intra}band contribution) in these formulas has been neglected as well.
For certain metals, e.g., K, Na, Be, Al, Mg, etc., this is a good approximation for the frequency range of interest.
But for other metals, e.g., Ag, the interband transitions or the core polarization ought to be taken into account.
Then, the core polarization replaces the unity on the right-hand side in Eq.~(\ref{eq 6}) (for $\epsilon'$) by $1+\epsilon_{inter}$.

For the sake of simplicity, we will assume that the dielectric matrix has no influence on the MN and can be characterized by
\be
 \label{eq 7}
  \epsilon'_m(\omega)=const\equiv\epsilon'_m, \qquad   \epsilon''_m(\omega)=0,
   \ee
i.e., the dielectric constant of the surrounding medium is assumed to be frequency independent.
However, it may happen in actual cases that the dielectric
medium is strongly absorptive at frequencies below the $\omega_{pl}$.
If that is the case, then the $\epsilon''_m$ is strongly dependent on
frequency and contributes to the attenuation of the oscillations.

Using Eqs.~(\ref{eq 6}) and (\ref{eq 7}), Eq.~(\ref{eq 4}) can be presented in the form
\be
 \label{eq 8}
  W(\omega)=\frac{1}{2}\epsilon'_m\sum_{j=1}^3
    |E_j^{(0)}|^2\;\omega\;{\rm Im}\;\alpha_{jj}\,(\omega),
     \ee
with
\be
 \label{eq 9}
  {\rm Im}\;\alpha_{jj}\,(\omega)=\left(\frac{V}{4\pi L_j}\right)
   \frac{\omega^3\xi_{j,m}\,\Gamma_{j}\,(\omega)} {(\omega^2-\omega^2_{j,{\rm sp}})^2+(\omega\Gamma_j(\omega))^2}
    \ee
is the imaginary part of the dynamic polarizability in Lorenzian form.\cite{G}
The designations introduced in Eq.~(\ref{eq 9}) have the following senses:
\be
 \label{eq 10}
  \xi_{j,m}=\frac{\epsilon'_m}{\epsilon'_m+L_{j}(1-\epsilon'_m)}
   \ee
is the dimensionless parameter,\cite{G}
\be
 \label{eq 11}
  \omega^2_{j,{\rm sp}}=\frac{L_{j}}{\epsilon'_m+L_{j}(1-\epsilon'_m)} \;\omega^2_{\rm pl}
   \ee
are the frequencies of the surface plasmon resonances, and
\be
 \label{eq 12}
  \Gamma_{j}(\omega)=\frac{4\pi L_{j}}
   {\epsilon'_m+L_{j}(1-\epsilon'_m)} \sigma'_{jj}(\omega)
    \ee
defines the damping rate, linewidth or, correspondingly, the decay time of the plasmon
resonance due to electron scattering both from the bulk and from the surfaces of the particle.
For understanding the decay mechanism of the electron plasma oscillations
the knowledge of the decay time is of central importance.

The $f$-sum rule ought to be fulfilled in these processes
\be
 \label{eq 13}
  \frac{2}{\pi}\int_0^\infty \;\omega\;{\rm Im}\;\alpha_{jj}\,(\omega)\;d\omega = N,
   \ee
where $N$ is the total number of electrons in the MN.

In the case of medium with $\epsilon_m\rightarrow 1$, Eqs.~(\ref{eq 8})--(\ref{eq 10}) are reduced merely to
\be
 \label{eq 14}
  \xi_{j,m}=1,
   \ee
\be
 \label{eq 15}
  \omega^2_{j,{\rm sp}}=L_{j}\;\omega^2_{\rm pl},
   \ee
\be
 \label{eq 16}
  \Gamma_{j}(\omega)=4\pi L_{j}\;\sigma'_{jj}(\omega).
   \ee
So, the decay time of the plasmon resonance is the electric conductivity of the MN
at the light frequency (optical conductivity) multiplied by a geometrical factor.

To study the dependence $\Gamma_j(\omega)$, it is necessary to find the real part of the conductivity tensor as a function of frequency.
There are different possibilities to calculate $\sigma'_{jj}(\omega)$ for the different frequency region. Below, we will apply the
kinetic equations approach. Benefit of this approach is that it permits one to study the effect of the particle shape on the
measured physical values. Second, it enables us to investigate the particles whose sizes are so small that the particle surface
start to play an important role.

Using this method, we have found earlier\cite{GT1} the general relation for complex conductivity tensor
\begin{eqnarray}
 \label{eq 17}
  \langle\sigma_{jj}^c({\bf r,\omega})\rangle & = &\sigma'_{jj} + i\sigma''_{jj} = \frac{3e^2 m^2}{(2\pi\hbar)^3}
   \frac{1}{\nu-i\omega} \nonumber \\&\times &
    \int d^3\upsilon\,\upsilon^2_{j}\,\delta(\upsilon^2-\upsilon^2_F)\,\Psi(q),
     \end{eqnarray}
where $e$ and $m$ are, respectively, the charge and mass of an electron, $\nu$ is the electron collision frequency,
$\upsilon_j$ is the $j$-th component of the electron velocity and $\upsilon_F$ refers to the electron velocity at the Fermi surface.
The complex $\Psi$ function entering in Eq.~(\ref{eq 17}), has the form
\be
 \label{eq 18}
  \Psi(q)=\Phi(q)-\frac{4}{q^2}\left(1+\frac{1}{q}\right)e^{-q},
   \ee
with
\be
 \label{eq 19}
  \Phi(q)=\frac{4}{3}-\frac{2}{q}+\frac{4}{q^3},   \qquad   q=\frac{2R}{\upsilon'}(\nu-i\omega),
   \ee
and $\upsilon'$ $(=\varsigma\upsilon)$ is a "deformed" electron velocity\cite{TG} with
the "deformation" coefficient $\varsigma_j=R/R_j$. The last summand in Eq.~(\ref{eq 18})
represents the oscillation part of the $\Psi$ function and the first one refers to its smooth part.

Further, we will restrict ourselves to the nanoparticles with a {\em spheroidal shape} only.
In this case, we have found that the components of the conductivity tensor for
light polarized along ($\|$) or across ($\bot$) the rotation axis of a spheroidal MN are
\be
 \label{eq 20}
  \sigma'_{\|\choose\bot}(\omega)=\frac{9}{4}\frac{ne^2}{m}{\rm Re}\left[\frac{1}{\nu-i\omega}\int\limits_0^{\pi/2}
   {\sin\theta\,\cos^2\theta \choose \frac{1}{2}\sin^3\theta}\Psi(\theta)\;d\theta\right]_{\upsilon=\upsilon_F},
    \ee
where $n$ is the electron concentration and $\theta$ is the angle between rotation axes of
the spheroid and direction of an electron velocity. Here and below, the upper (lower) symbol in the
parentheses on the left-hand side of Eq.~(\ref{eq 20}) corresponds to the upper (lower) expression
in the parentheses on the right-hand side of this equation.
The subscript $\upsilon=\upsilon_F$ means that the electron velocity in the final expressions should be taken on the Fermi surface.
The $\Psi$ function in Eq.~(\ref{eq 20}) varies now with the angle $\theta$ because the $q$
for a spheroidal particle becomes dependent on the angle $\theta$, namely
\be
 \label{eq 21}
  q=\frac{2}{\upsilon_F}
   \frac{\nu-i\omega}{\sqrt{\frac{\cos^2\theta}{R^2_{\|}}+\frac{\sin^2\theta}{R^2_{\bot}}}}\equiv q(\theta),
    \ee
where $R_{\|}$ and $R_{\bot}$ are the spheroid semiaxes directed along and across the spheroid rotation axis,
respectively.\cite{FN1}
The semiaxial ratio $R_{\bot}/R_{\|}$ is a measure of the shape of MN.
The semiaxes are connected to the radius of sphere $R$ of an equivalent volume through the relation $R^3={R_{\|}}R^2_{\bot}$.

The species of Eq.~(\ref{eq 21}) for $q$ is governed by the "deformed" electron velocity
entering into Eq.~(\ref{eq 19}), which in the case of a spheroidal MN takes the form
\be
 \label{eq 22}
  \upsilon'=\upsilon R\sqrt{\left(\frac{\sin\theta}{R_{\bot}}\right)^2+
   \left(\frac{\cos\theta}{R_{\|}}\right)^2}\equiv\upsilon'(\theta),
    \ee
where $\upsilon_{\|}=\upsilon\cos\theta$ and $\upsilon_{\bot}=\upsilon\sin\theta$ are
the velocity components along and across to the spheroid rotation axis, respectively.
In the case of MN with a spherical shape, $R_{\|}=R_{\bot}\equiv R$, $\upsilon'= \upsilon$,
and the $\Psi$ function ceases to depend on the angle $\theta$.

Below, we will consider some particular cases that enable us to derive the explicit analytical expressions for $\Gamma(\omega)$.

\section{Particular Cases}

Let us introduce the frequency of electron oscillations between particle walls as
\be
 \label{eq 23}
   \nu_s=\frac{\upsilon_F}{2R}.
    \ee
Depending on sizes of MN, its shape and temperature, the variety of relations
between frequencies $\nu_s$, $\nu$ and $\omega_{\rm sp}$ can be achieved.
For example, for the Na nanoparticle with the radius of $R < 2~\AA$, $\nu_s\simeq\omega_{\rm sp}$.
On the other hand, with $R > 126~\AA$, the electron oscillation frequency becomes $\nu_s < \nu$,
where $\nu\simeq 4.24\cdot 10^{13}$~s$^{-1}$ is estimated for the Na at 300$^0$~K.
This leads to a different expressions for $\sigma(\omega)$, which can be used
in Eqs.~(\ref{eq 12}) and (\ref{eq 16}) for calculation of the plasmon linewidth.

\subsection{Plasmon linewidth of a spheroidal MN with an account for the bulk damping}

The components of the electric conductivity can be represented in terms of an expansion in a power series of $\nu/\nu_s$.
Then from Eq.~(\ref{eq 20}) within the frequency region $\omega\gg\nu_s\gg\nu$, one gets
\be
 \label{eq 24}
  \sigma'_{\|\choose\bot}(\omega)\simeq\frac{1}{4\pi}\left(\frac{\omega_{\rm pl}}{\omega}\right)^2
   \left(\nu+\frac{9}{2}\nu_s I_{\|\choose\bot}+\cdots\right).
    \ee

In the case of $\omega\gg\nu_s$, but $\nu\gg\nu_s$, we have
\be
 \label{eq 25}
  \sigma'_{\|\choose\bot}(\omega)\simeq\frac{\omega^2_{\rm pl}}{4\pi}\left(\frac{\nu}{\nu^2+\omega^2}-
   \frac{9}{2}\nu_s\frac{\nu^2-\omega^2}{(\nu^2+\omega^2)^2}I_{\|\choose\bot}+\cdots\right).
    \ee
The factors $I_{\|}$ and $I_{\bot}$ entering in Eqs.~(\ref{eq 24}) and (\ref{eq 25}) depend
only on the spheroid axial ratio $x$ ($=R_{\bot}/R_{\|}$), and have the form
\be
 \label{eq 26}
  I_{\|}=
   \frac{x^{2/3}}{8}\left(2+\frac{1}{x^2-1}\right)-
    \frac{\ln\left|x+\sqrt{x^2-1}\right|}{8x^{1/3}(x^2-1)^{3/2}},
     \ee
and
\begin{eqnarray}
 \label{eq 27}
  I_{\bot}&=&
   \frac{x^{2/3}}{16}\left(2-\frac{1}{x^2-1}\right)+\frac{1}{4x^{1/3}\sqrt{x^2-1}}\nonumber \\&\times&
    \left(1+\frac{1}{4(x^2-1)}\right)\ln\left|x+\sqrt{x^2-1}\right|.
     \end{eqnarray}
For nanoparticles with a spherical shape ($x\rightarrow 1$), $I_{\|}=I_{\bot}=1/3$.
In the case of the particles with a prolate shape ($x<1$), one should perform in Eqs.~(\ref{eq 26})
and (\ref{eq 27}) the following replacement: $\ln\left|x+\sqrt{x^2-1}\right|\rightarrow i\arcsin{\sqrt{1-x^2}}$.
For strongly oblate ($x\gg 1$) or prolate ($x\ll 1$) MNs, Eqs.~(\ref{eq 26}) and (\ref{eq 27}) reduce to the form
\be
 \label{eq 28}
  I_{{\|\choose\bot},{\rm ob}}\simeq\frac{x^{2/3}}{4} {1\choose 1/2},
   \quad I_{{\|\choose\bot},{\rm prol}}\simeq\frac{\pi}{16}\frac{1}{x^{1/3}}{1\choose 3/2},
    \ee
respectively.

Therefore, the line-width of the plasmon resonances for two light polarizations
($\|$ and $\bot$ to the spheroid rotation axis) takes the form
\be
 \label{eq 29}
  \Gamma_{\|\choose\bot}(\omega) = \frac{\left(\frac{\omega_{\rm pl}}{\omega}\right)^2 L_{\|\choose\bot}}  {\epsilon'_m+L_{\|\choose\bot}(1-\epsilon'_m)} \left(\nu+\frac{9}{2}\nu_s I_{\|\choose\bot}+\cdots\right),
    \ee
provided that $\omega\gg\nu_s\gg\nu$, and
\begin{eqnarray}
 \label{eq 30}
  &&\Gamma_{\|\choose\bot}(\omega)=\frac{\omega^2_{\rm pl} L_{\|\choose\bot}}
   {\epsilon'_m+L_{\|\choose\bot}(1-\epsilon'_m)} \nonumber \\&\times& \left(\frac{\nu}{\nu^2+\omega^2}-
    \frac{9}{2}\nu_s\frac{\nu^2-\omega^2}{(\nu^2+\omega^2)^2}I_{\|\choose\bot}+\cdots\right),
     \end{eqnarray}
provided that $\omega\gg\nu_s$ and $\nu_s\ll\nu$. The factors $L_{\|}$ and $L_{\bot}$ in Eqs.~(\ref{eq 29})
and (\ref{eq 30}) are the longitudinal and transverse components of the geometrical factor, respectively.

\subsection{Plasmon linewidth of a spheroidal MN when the bulk damping is neglected}
\subsubsection{Plasmon linewidth of a spheroidal MN in HF limit}

The components of conductivity tensor for a spheroidal MN in the highfrequency
limit ($\omega\gg\nu_s$) and $\nu_s\gg\nu$, can be represented as\cite{GT3}
\be
 \label{eq 31}
  \sigma'_{\|\choose\bot}(\omega)=\frac{9}{32\pi}\left(\frac{\omega_{\rm pl}}{\omega}\right)^2
   \frac{\upsilon_F}{R_{\bot}}{\eta(e_p) \choose \rho(e_p)},
    \ee
where $R_{\bot} (= R x^{1/3})$ is spheroid semiaxis directed across to the spheroid rotation axis,
and $\eta(e_p)$ and $\rho(e_p)$ are smooth functions dependent only on the spheroid
eccentricity $e_p=\sqrt{1-x^2}$ (a prolate spheroid), or $e_p= \sqrt{x^2-1}$ (an oblate one):
\begin{widetext}
 \be
  \label{eq 32}
    \eta(e_p)=\left\{
     \begin{array}{ll}
      -\frac{1}{4e_p^2}\left(1-2e^2_p\right)\sqrt{1-e^2_p}+
       \frac{1}{4e_p^3}\arcsin{e_p}, & \textrm{for a prolate spheroid} \\
        \frac{1}{4e_p^2}\left(1+2e^2_p\right)\sqrt{1+e^2_p}-
         \frac{1}{4e_p^3}\ln{\left(e_p+\sqrt{1+e^2_p}\right)}, & \textrm{for an oblate one}
          \end{array}
           \right. ,
            \ee
\be
 \label{eq 33}
   \rho(e_p)=\left\{
    \begin{array}{ll}
     \frac{1}{8e^2_p}(1+2e^2_p)\sqrt{1-e^2_p}-\frac{1}{8e^3_p}\left(1-
      4e^2_p\right)\arcsin{e_p}, & \textrm{for a prolate spheroid} \\
      -\frac{1}{8e^2_p}(1-2e^2_p)\sqrt{1+e^2_p}+\frac{1}{8e^3_p}\left(1+
        4e^2_p\right)\ln{\left(e_p+\sqrt{1+e^2_p}\right)},
        & \textrm{for an oblate one}
         \end{array}
          \right. .
           \ee
\end{widetext}

Eq.~(\ref{eq 12}) with accounting for Eq.~(\ref{eq 31}), takes the form
\be
 \label{eq 34}
  \Gamma_{\|\choose\bot}(\omega)=\frac{\frac{9}{8}
   \left(\frac{\omega_{\rm pl}}{\omega}\right)^2\frac{\upsilon_F}{R_{\bot}}}
    {\epsilon_m+L_{\|\choose\bot}(1-\epsilon_m)}
     L_{\|\choose\bot} {\eta(e_p) \choose \rho(e_p)}.
      \ee

Using Eq.~(\ref{eq 11}) for the plasmon resonance frequencies ($\omega=\omega_{j,\rm sp}, j=\|,\bot$),
Eq.~(\ref{eq 34}) can be reduced to the form
\be
 \label{eq 35}
  \Gamma_{\|\choose\bot}(x)=\frac{9}{8}\frac{\upsilon_F}{Rx^{1/3}} {\eta(x) \choose \rho(x)},
   \ee
where the product $R\,x^{1/3}$ represent the $R_{\bot}$. In this case the linewidth depends
on the shape of MN solely through the functions $\eta(x)$ and $\rho(x)$.

One can use the asymptotic expressions for functions $\eta(x)$ and $\rho(x)$
in the cases of both the extremely small or the large axial ratio:
\be
 \label{eq 36}
  \begin{array}{ll}
   &\eta(x)\simeq\left\{
    \begin{array}{ll}
     \pi/8+3\pi x^2/16, & x\ll 1 \\ \\
      x/2+1/(4x), & x\gg 1
       \end{array}
        \right.,
\\ \\
&\rho(x)\simeq\left\{
 \begin{array}{ll}
  3\pi/16+\pi x^2/32, & x\ll 1 \\ \\
   x/4+(-1+4\ln{2x})/(8x), & x\gg 1
    \end{array}
     \right..
     \end{array}
      \ee

If one consider the nanowires and nanorods, which can be reasonable approximated as prolate spheroids, then one
can put $\eta\simeq\pi/8$ and $\rho\simeq 3\pi/16$ with a sufficient degree of accuracy.
The depolarization coefficients in this case ($x\ll 1$) look as
\be
 \label{eq 37}
  L_{\|}(x)\simeq x^2 \left[\ln{(2/x-x/4)}-1\right], \quad  L_{\bot}(x)=[1-L_{\|}(x)]/2.
   \ee

In the case of MNs with a spherical shape $\eta=\rho=2/3$.

\subsubsection{Plasmon linewidth of a spherical MN}

In the case of MN with a spherical shape, one can put the depolarization factor equal
to $L_{\|}=L_{\bot}=1/3$ in Eqs.~(\ref{eq 29}), (\ref{eq 30}), and (\ref{eq 34}).
In the most common case, Eq.~(\ref{eq 12}) then reduces to the form
\be
 \label{eq 38}
  \Gamma_{\rm sph}(\omega)=\frac{4\pi}{2\epsilon_m+1} \sigma'_{\rm sph}(\omega).
   \ee

To calculate the $\Gamma_{\rm sph}(\omega)$, one need only the function $\sigma'_{\rm sph}(\omega)$.
Let us choose for illustration the case $\nu\ll\nu_s$.
The kinetic approach in this case gives the expression\cite{GT1}
\be
 \label{eq 39}
  \sigma'_{\rm sph}\simeq\frac{3}{8\pi}\nu_s\frac{\omega^2_{\rm pl}}{\omega^2}\left[1-
   \frac{2\nu_s}{\omega}\sin\frac{\omega}{\nu_s}+\frac{2\nu^2_s}{\omega^2}\left(1-
    \cos\frac{\omega}{\nu_s}\right)\right].
     \ee
Substituting Eq.~(\ref{eq 39}) into Eq.~(\ref{eq 38}), we obtain
\begin{eqnarray}
 \label{eq 40}
  \Gamma(\omega)&\simeq &\frac{\upsilon_F}{4R}
   \left(\frac{\omega_{\rm pl}}{\omega}\right)^2 \frac{3}{2\epsilon'_m+1}\nonumber \\&\times&
   \left[1-\frac{2\nu_s}{\omega}\sin\frac{\omega}{\nu_s}+\frac{2\nu^2_s}{\omega^2}\left(1-
    \cos\frac{\omega}{\nu_s}\right)\right].
    \end{eqnarray}
Taking into account only the first term in (\ref{eq 40}), we recover at $\epsilon'_m=1$
the well-known\cite{KV,KK,YB,MWJ} $1/R$ dependence of $\Gamma$
\be
 \label{eq 41}
  \Gamma_0(\omega, R)=\frac{1}{4}\frac{\upsilon_F}{R}
   \left(\frac{\omega_{\rm pl}}{\omega}\right)^2.
    \ee
As seen from Eqs.~(\ref{eq 40}) and (\ref{eq 41}), the lifetime (which can be estimated as $1/\tau=\Gamma$) of an excitation
in the MN depends not only on the nanoparticle radius, but also on the frequency (at which a given excitement is reasonable).
For frequency that corresponds to the excitation of a surface plasmon in MN in a vacuum, $\omega=\omega_{\rm pl}/\sqrt{3}$,
the following relation can be obtained from Eq.~(\ref{eq 41}) in energy units:
\be
 \label{eq 42}
  \Gamma_0^{\rm sp}(R)=\frac{3}{4}\hbar\frac{\upsilon_F}{R}.
   \ee

The same result can be derived for spherical MNs using the relation connecting
a plasmon linewidth directly with a dielectric function\cite{KV,CJC}
\be
 \label{eq 43}
  \left.\Gamma\simeq\frac{2\epsilon''(\omega)}{|d\epsilon'/d\omega|}\right|_{\omega=\omega_{\rm sp}}.
   \ee
To make sure in that, it is enough to rewrite Eq.~(\ref{eq 43}) using Eq.~(\ref{eq 6}) in the form
\be
 \label{eq 44}
  \Gamma\simeq 4\pi\left(\frac{\omega}{\omega_{\rm pl}}\right)^2\sigma'(\omega)\; |_{\omega=\omega_{\rm sp}}.
   \ee
Then substituting Eqs.~(\ref{eq 31}) with $\eta_e^H=\rho_H=2/3$ into Eq.~(\ref{eq 44}), we come to the result of Eq.~(\ref{eq 42}).

The oscillating terms in Eq.~(\ref{eq 40}) give rise to the oscillation of $\Gamma$ around of $\Gamma_0$ as a function
of both the particle radius and the frequency. They can be represented at the frequency of a surface plasmon as follows:
\begin{eqnarray}
 \label{eq 45}
  &&\Gamma_{\rm osc}^{\rm sp}(R)\simeq \frac{3\sqrt{3}}{4}\frac{\hbar}{\omega_{\rm pl}}\left(\frac{\upsilon_F}{R}
   \right)^2 \frac{3}{2\epsilon'_m+1}\nonumber \\&\times & \left[-\sin\frac{2R\,\omega_{\rm pl}}{\sqrt{3}\;\upsilon_F}+
    \frac{\sqrt{3}\;\upsilon_F}{2R\,\omega_{\rm pl}}\left(1-\cos\frac{2R\,\omega_{\rm pl}}{\sqrt{3}\;\upsilon_F}\right)\right].
     \end{eqnarray}
The amplitude and period of oscillations can be evaluated by means of the following relations
\be
 \label{eq 46}
  \Gamma^{\rm max}_{\rm osc}\simeq \frac{9\sqrt{3}\,\hbar\upsilon_F^2}{4\omega_{\rm pl}R^2(2\epsilon'_m+1)},
   \quad T=\frac{\sqrt{3}\,\pi\upsilon_F}{\omega_{\rm pl}R},
    \ee
respectively.

\section{Discussions of Results}
\subsection{Plasmon linewidth in a general case}

In order to study the significance of the oscillatory behavior in more general situations, it is necessary
to perform the numerical calculations in Eq.~(\ref{eq 12}) with the use of a general expression for the conductivity
tensor given by Eq.~(\ref{eq 20}). By analogy with Eqs.~(\ref{eq 40}), the expression (\ref{eq 12})
with $\sigma'_{jj}(\omega)$, presented by Eq.~(\ref{eq 20}), can be separated as well in two terms describing
the smooth and oscillatory parts of the plasmon linewidth: $\Gamma_0+\Gamma_{\rm osc}$.
To get the smooth part, we restrict ourselves only to the first term in Eq.~(\ref{eq 18}).
To have oscillatory part, we retain only the last term in Eq.~(\ref{eq 18}).

\noindent\includegraphics[width=8.6cm]{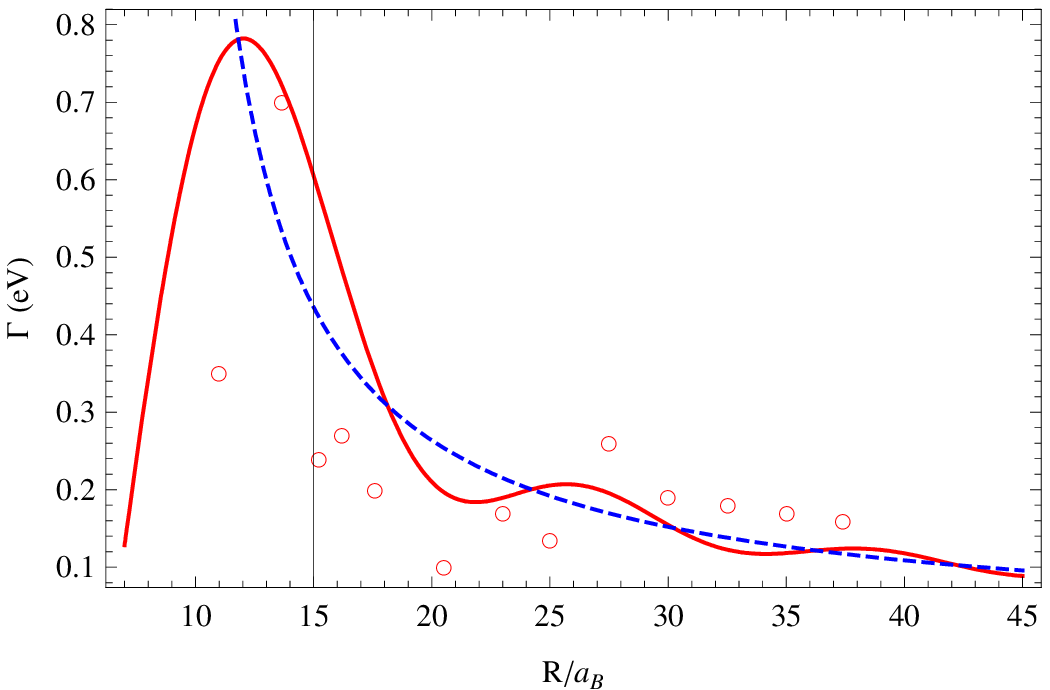}
\vskip-1mm\noindent{\footnotesize FIG.~1. (Color online) Linewidth of the surface plasmon
resonance vs radius (in units of Bohr radius $a_B\simeq 0.53 \AA$) of the spherical Na nanoparticles
embedded in medium with $\epsilon_m=1$.
The hollow rings correspond to the TDLDA calculation data taken from Ref.~\onlinecite{MWJ}.
The smooth term $\Gamma_0(R)$ is given by the dashed line and the solid line corresponds to the sum of both terms
in Eq.~(\ref{eq 18}).}
\vskip10pt

Fig.~1 shows the full linewidth $\Gamma=\Gamma_0+\Gamma_{\rm osc}$ of the plasmon resonance as a function
of a particle radius. Here and below, the results were obtained by numerical evaluating of Eqs.~(\ref{eq 12})
and (\ref{eq 20}) for the Na nanoparticle with the parameters\cite{K}: $n_e\simeq 2.65\times 10^{22}$~cm$^{-3}$,
$\upsilon_F=1.07\times 10^{8}$~cm/s and $\omega_{\rm pl}=9.18\times 10^{15}$~s$^{-1}$.
We have taken  $\nu_{300^0{\rm C}}\simeq 4.2\times 10^{13}$ s$^{-1}$, which, is assumed to be an
appropriate value for the Na nanoparticle.
Knowing only $n_e$, the parameters $\omega_{\rm pl}$ and $\upsilon_F$ can be estimated therewith by means of the formulas:
\be
 \label{eq 47}
  \omega_{\rm pl}=\sqrt{4\pi n_e e^2/m}, \qquad \upsilon_F=\frac{2\pi\hbar}{m}\left(n_e\frac{3}{8\pi} \right)^{1/3}.
   \ee

Mostly, for the kinetics, the following inequality could be met:
\be
 \label{eq 48}
  R\gg R_{\rm dB}=\frac{2\pi\hbar}{m\upsilon_F}.
   \ee
For the Na particles, e.g., $R\gg R_{\rm dB}\simeq 7\AA$.
Since we can not apply the kinetic method to the range of $R$, where the quantum effects (such as, e.g.,
the Landau damping) play an important role, we have restricted ourselves to some minimal
value of $R_{\rm min}$, from which the $R$ should be measured.
To fit our calculations to the TDLDA calculations
for the Na cluster presented in Ref.~\onlinecite{MWJ}, we choose though, $R_{\rm min}=R_{\rm dB}/2$.

As can be seen in Fig.~1, the calculated smooth component of the SP linewidth $\Gamma_0(R)$
is inversely proportional to the radius of the nanoparticle.
The resonance peaks are sharper for larger particle radii and tend to broaden for lower particle radiuses.
The oscillating terms represent an important correction to $\Gamma_0(R)$, especially
at small particle radii. Additionally, one can see that the reduction of the MN radius up to $R/a_B\simeq 12$ leads
to the increase of $\Gamma$ to the maximal value around of 0.78~eV (in the time domain it corresponds to the minimal SP
dephasing time of $\tau_{\rm min, sp}\simeq 1.7$~fs).
In other words, the dephasing rate of spherical MHs increases for larger particles.
A further reduction of $R$ to $R/a_B\simeq 8$ causes the decrease of $\Gamma$ to values as low as 0.3~eV.
Such a behavior of $\Gamma(R)$ (without oscillations) resembles qualitatively the measured one for the Ag nanoparticles.\cite{BHS}

Our result for the Na nanoparticles mainly agrees with a similar results obtained in Refs.~\onlinecite{MWJ},\onlinecite{WMW}.
Experiments on alkaline clusters with a diameter in the range of $10-50~\AA$ in vacuum\cite{BCL}
yield a linewidth of the order of $\Gamma\sim 1$~eV. Our calculated value is slightly smaller,
but of the same order of magnitude as the experimental one.

\subsection{Particle shape effect}

The next interesting question is the effect of a particle shape on the plasmon relaxation dynamics.
The shape factor of the ellipsoidal particles can be parameterized by the $R_{\bot}/R_{\|}\equiv x$ ratio.
Oblate ellipsoids correspond to the case of $x > 1$ (pancake shaped), spherical particles --
to $x =1$, and prolate ellipsoids -- to $x < 1$ (cigar shaped).
As already was outlined,\cite{BH,KV,G} any change of the nanoparticle shape from a sphere and thus introducing
of an anisotropy, results in the splitting of the surface plasmon resonance into two modes: a transverse one
(perpendicular to the spheroid axis of revolution) and a longitudinal one (parallel to this axis).

In the investigated size regime, the frequency of the resonance absorption is rigidly determined by the shape
of MN.\cite{KV,GT3} With increasing particle {\em prolateness}, the $\|$-component of the surface plasmon
resonance peak (at $\omega<\omega_{sp}$, recall that $\omega_{\rm sp}=\omega_{\rm pl}/\sqrt{3}$ refers to the SP
frequency for MN with a spherical shape) shifts to the red side, whereas the $\bot$-component (at $\omega>\omega_{sp}$)
of the plasmon peak shifts to the blue side of the spectrum.\cite{G} For more and more {\em oblate}-shaped MNs,
on the contrary, the $\bot$-component of the peak (at $\omega<\omega_{sp}$) shifts to the red side,
and the $\|$-component (at $\omega>\omega_{sp}$) shifts to the blue side of the spectrum.

\noindent\includegraphics[width=8.6cm]{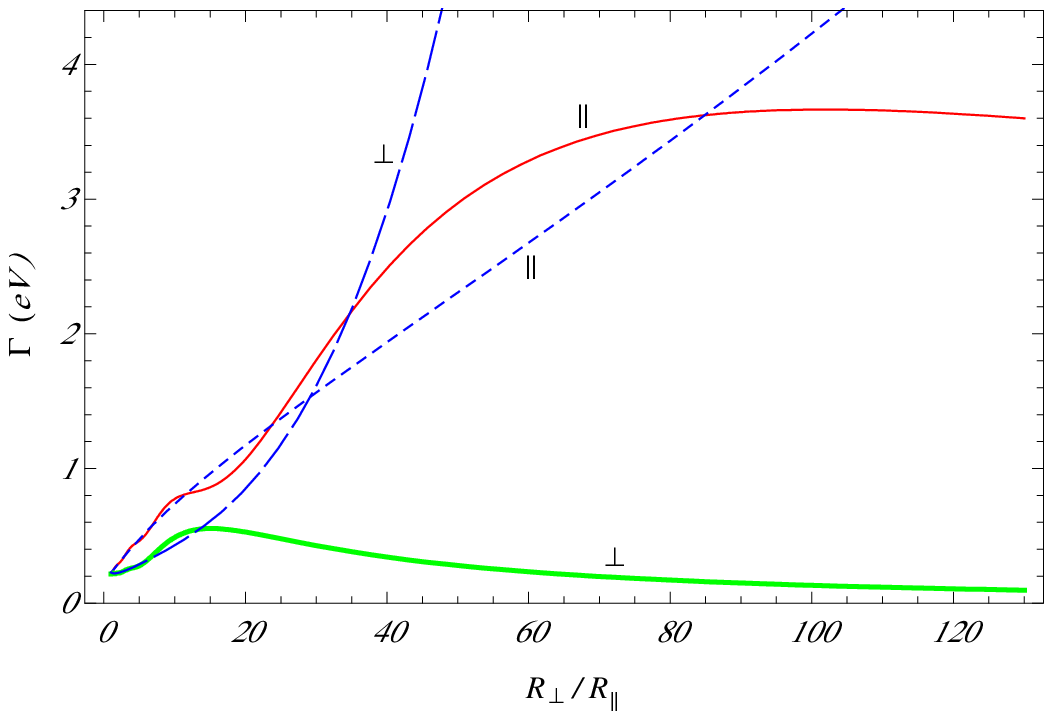}
\vskip-1mm\noindent{\footnotesize FIG.~2. (Color online) Linewidth of the components $\Gamma_{||}$ (thin line)
and $\Gamma_{\bot}$ (thick line) of the surface plasmon resonances $\omega_{\|,pl}$ and $\omega_{\bot,pl}$
vs axial ratio $R_{\bot}/R_{||}$ for the Na nanoparticles with the radius $20 \AA$ embedded in the medium
with $\epsilon_m=1$. The short-dashed ($\Gamma_{0,\|}$) and long-dashed ($\Gamma_{0,\bot}$) lines
correspond to the calculations of Eq.~(\ref{eq 18}) without account for the oscillations terms. }
\vskip10pt
\noindent\includegraphics[width=8.6cm]{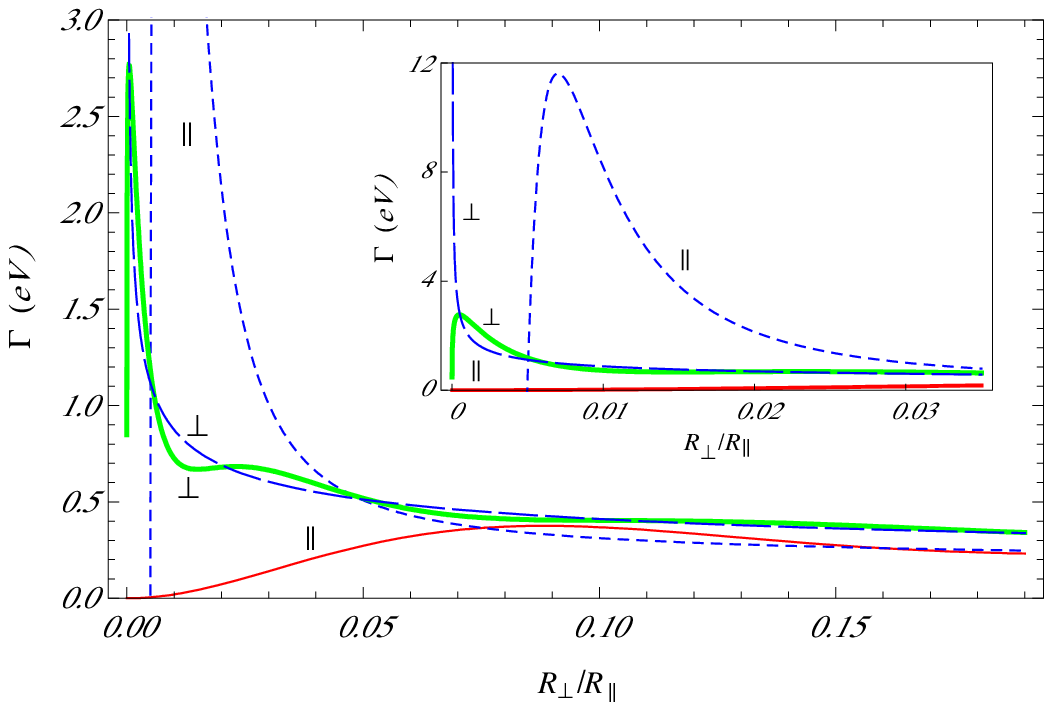}
\vskip-1mm\noindent{\footnotesize FIG.~3. (Color online)
The same as in Fig.~2, for the prolate MNs.
Inset shows the $\Gamma(x)$ at small $x$ ratio.}
\vskip10pt

We investigate also the behavior of the plasmon resonance linewidth of a MN with the variation of MN shape.
In Figs.~2 and 3 the dependence of the two components of the linewidth on the axis ratio $x$
of the oblate and prolate Na particles, respectively, are presented for two SP resonance frequencies $\omega_{\|, sp}$
and $\omega_{\bot, sp}$. The radius of the sphere of an equivalent particle volume was taken to be $R=20\AA$.
It is supposed that the Na particle is placed in vacuum.

\begin{table*}
\caption{\label{tab:table3}The extremal values of the linewidth components
($\Gamma_{\|}$ and $\Gamma_{\bot}$) for the Na nanoparticles with different radii.}
\begin{ruledtabular}
\begin{tabular}{ccccccccccccc}
 &\multicolumn{6}{c}{Oblate Na $(x>1)$}   $|$  &\multicolumn{6}{c}{Prolate Na $(x<1)$}\\
 $R$(\AA)&$\Gamma_{\|,{\rm max}}$(eV)& $x$ &$\Gamma_{\bot,{\rm max}}$(eV)&$x$
&$\Gamma_{\bot,{\rm min}}$(eV)& $x$ & $\Gamma_{\|,{\rm max}}$(eV)& $x$ & $\Gamma_{\bot,{\rm max}}$(eV)&$ x $
&$\Gamma_{\bot,{\rm min}}$(eV)& $x$ \\ \hline
 20&$3.67$&$102$ &$0.555$&$15$ &$0.2185$&$1.5$ &$0.375$&$0.088$ &$2.765$&$0.0005$  &$0.67$&$0.015$ \\
 20&&&0.2195\footnotemark[1]&1.28&0.2165&1.03&&&$0.684\footnotemark[1]$&$0.023$&&\\
 100&$3.69$&$823$&$0.31$&$51$ &$0.076$&$1.5$ &$0.14$&$0.028$ &$2.765$&$0.8\times 10^{-5}$ &$0.67$&$0.00023$\\
 100&&&&&&&&&$0.684\footnotemark[1]$&$0.00035$
\end{tabular}
\end{ruledtabular}
\footnotetext[1]{This is for the second maximum of $\Gamma_{\bot}$.}
\end{table*}

The numerical calculations were carried out using Eqs.~(\ref{eq 12}) and (\ref{eq 20}) with the same
numerical parameters as was above-mentioned. The smooth parts of the damping $\Gamma_{0,\|}$ (short-dashed lines)
and $\Gamma_{0,\bot}$ (long-dashed lines) correspond to the case when the oscillatory terms in Eq.~(\ref{eq 18})
was neglected. The inset in Fig.~3 shows the behavior of $\Gamma$ at small axes ratio.
One can see that the smooth part of the damping $\Gamma_{0,\|}$ has the maximum too.

As is seen from Figs.~2 and 3, both components of the linewidth oscillate around the dashed lines
with the axial ratio variation.
The period of these oscillations depends on the particle volume and enhances with the $x$ growth.
The amplitude of oscillations enhances also with volume contraction. However, the oscillations
of $\Gamma$ are disappeared for some axis ratio $x$, when the $\Gamma$ reaches maximum (Fig.~2).

 The peaks of $\Gamma$ mean that among a variety of shapes of MN there exist such, for which the
 surface plasmon lifetime is minimal. Vice versa, the minimums of $\Gamma$ means that among a variety
 of shapes of MN there exists such, for which the surface plasmon lifetime is maximal.

The maximal and the minimal values of two linewidth components, deduced from Figs.~2 and 3,
are tabulated in Table~I for both the oblate and the prolate Na nanoparticle with two different radii.
From Table~I we observe that with bulking of an {\em oblate} MN (equivalent to an increase its radius),
the maximum of both components of $\Gamma$ shifts to the side of the larger values of the $R_{\bot}/R_{\|}$;
the absolute value of the $\Gamma_{\bot}$-component tends to diminish and the $\Gamma_{\|}$-component retains approximately the same.
The minimal linewidth $\Gamma_{\rm min}$ was detected only for the transverse component of the $\Gamma$.
With an increase of $R$, the location of the $\Gamma_{\rm min}$ does not change, but its intensity is substantially reduced.

 In the case of {\em prolate} Na particles with bulking of a MN, the maximum of both components of $\Gamma$,
 on the contrary, shifts toward the side of smaller values of the ratio $R_{\bot}/R_{\|}$; the absolute value of $\Gamma$
 is reduced for the $\Gamma_{\|}$-components, but for the $\Gamma_{\bot}$-components retains roughly the same.
 The minimal linewidth $\Gamma_{\rm min}$, was detected also only for the transverse component of the $\Gamma$.
The magnitude of the $\Gamma_{\rm min}$ does not change with growing of a particle size,
but its location is drastically shifted toward the side of small $x$.

In general, the resonance plasmon damping in the {\em oblate} Na nanoparticle
was found stronger along the spheroid revolution axis than the one across this axis.
For the {\em prolate} Na nanoparticle, on the contrary, the damping along the revolution axis
was weaker than the one across this axis. This result holds regardless of whether
the photo-excitation is close to the surface plasmon resonance or far from it.

If the MN is embedded in the dielectric media with $\epsilon_m > 1$,
then the environment effect ought to be taken into account.

\subsection{Environment effect}

Because the effect of an electric field on the embedded nanoparticles becomes weaker in a dielectric media proportionally
to its refractive index, the environment effect plays, additionally, an important role. The possibility to measure
the optical effects in any medium plays an important role for numerous applications in which the nanoparticle plasmon
is used as an optical sensor ($\partial\Gamma/\partial\epsilon_m$) for its dielectric surrounding.\cite{SFW}
Persson have studied this problem theoretically for small silver particles embedded in various matrixes.\cite{P}
The spectral peculiarities of an environment effect recently were investigated for the Ag and Au nanoparticles,
for instance, in works~[\onlinecite{ML,KCZ,TPF}].
Below, we will present the result of our calculation of the linewidth for the spherical Na nanoparticle
embedded in a different dielectric media.

\noindent\includegraphics[width=8.6cm]{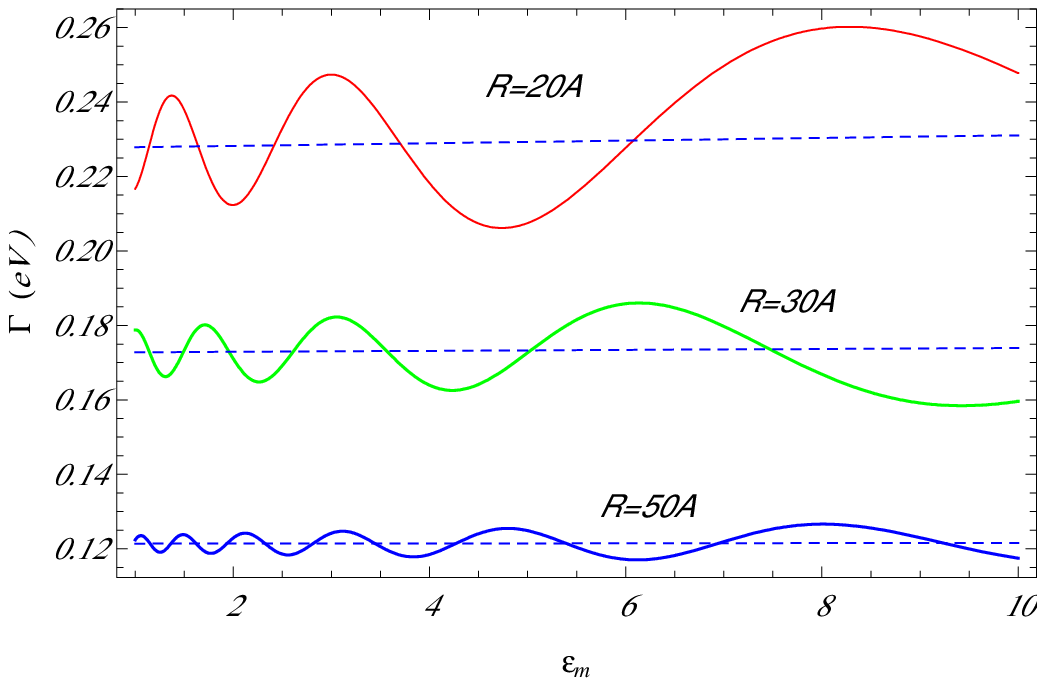}
\vskip-1mm\noindent{\footnotesize FIG.~4. (Color online) The total (solid lines) and smooth
(dashed lines) linewidth of the surface plasmon resonance vs dielectric constant of an environment
media for the Na nanoparticles with the radii of 20, 30, and $50 \AA$.}
\vskip10pt

The resonance plasmon linewidth, is plotted in Figure~4 against the dielectric constant of various
dielectric environments for the Na nanoparticles with three different radii. The numerical calculations
were performed for a plasmon resonance frequency $\omega=\omega_{\rm pl}/\sqrt{2\epsilon'_m+1}$,
with the use of Eqs.~(\ref{eq 12}) and (\ref{eq 20}).
As one can see in Fig.~4, the total linewidth of the surface plasmon resonance slightly rises
and oscillates round its smooth part when the dielectric constant of the environment is increased.
The oscillations are well pronounced for the Na nanoparticles with the small radii and disappeared for NP
with a larger radii. In experiments for MNs with small sizes (where the bulk contribution to the $\Gamma$
is decreased),\cite{HNF} only the linewidth reduction with $\epsilon_m$ was fixed.

Another results are as follows. With an increase of the particle radius, the linewidth of SPR
appreciably falls and oscillates around an constant value in higher dielectric constant environment.
The magnitude of these oscillations is the greater the smaller particle is and enhances markedly with $\epsilon_m$.

The resonance peak position gets shifted with changing the dielectric function of the surrounding medium.
The spectral direction of a shift depends on numbers of factors, well studied in earlier publications.\cite{BH,KV,HNF,ML,KCZ,TPF}

\subsection{Frequency dependence of $\Gamma$}

Let us study how the linewidth of a surface plasmon resonance changes with the deviating
of an incident light frequency from the resonance one.

Fig.~5, depicts the behavior of a plasmon resonance linewidth vs the deviation of a light frequency
from plasmon resonance one for the prolate Na nanoparticle with $R_{\bot}/R_{\|}=1/10$.
Because there is a direct connection between the plasmon resonance frequency $\omega_{sp}$ and
the axes ratio $R_{\bot}/R_{\|}$, the dependence of $\Gamma$ on the frequency can be treated as
well as the dependence of $\Gamma$ on the shape of spheroid.

The numerical calculation were carried out, as was done above, using Eqs.~(\ref{eq 12}), (\ref{eq 20})
and the same numerical parameters. The frequency scale is normalized to a plasmon resonance frequency
$\omega_{\rm sp}=\omega_{\rm pl}/\sqrt{3}\equiv\Omega$ for a spherical MN embedded in vacuum.

Most remarkable (see Fig.~5) is the strong damping of plasmon oscillations at a frequencies $\omega$
much lower than the $\omega_{\rm sp}$. This result is due only to the surface electron scattering.
Our calculation of the plasmon resonance
linewidth vs resonance energy qualitatively correlate with the its behavior measured for single
Au nanospheres.\cite{SFW} But a similar calculation for the Ag nanoparticles with account for the influence
of the bulk dielectric properties of the NPs on the damping process\cite{BHS} gives, on the contrary,
the increase of $\Gamma$ with the energy of a plasmon resonance. This may be due to the fact that
the bulk damping exceeds the surface one for MNs with the great radii.

As can be seen from Fig.~5, the $\bot$-component of the $\Gamma$ exceeds the $\|$-component one within a broad frequency range.
The inset shows that the damping is considerably decreased as the radius of the spherical particle increases.
Additionally, the oscillations of the $\Gamma$ gradually disappear as a particle becomes larger.

\noindent\includegraphics[width=8.6cm]{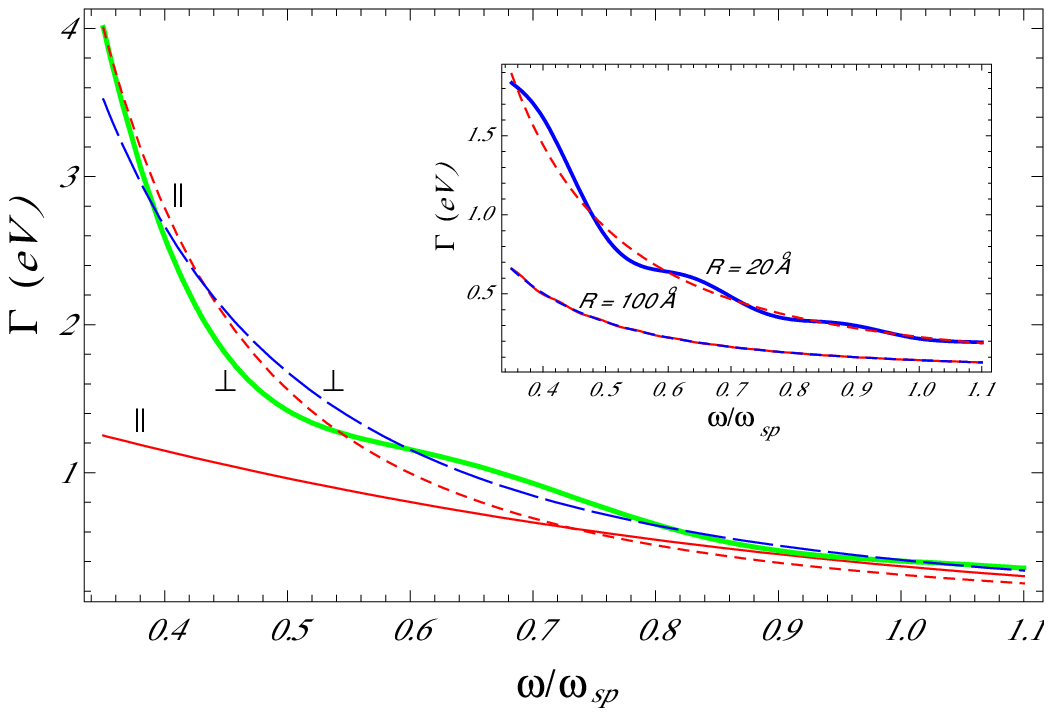}
\vskip-1mm\noindent{\footnotesize FIG.~5. (Color online) Linewidth of the components $\Gamma_{||}$
(thin line) and $\Gamma_{\bot}$ (thick line) of the surface plasmon resonances for {\em prolate}
Na particles with $x=1/10$, embedded in the medium with $\epsilon_m=1$, as a function of frequency in the vicinity of the
$\omega_{\rm sp}$. The short-dashed ($\Gamma_{0,\|}$) and long-dashed ($\Gamma_{0,\bot}$)
lines correspond to the calculations with accounting for only the smooth term in Eq.~(\ref{eq 18}).
Inset shows the same dependence for the {\em spherical} Na particles with $R=20\AA$ and $100\AA$. }
\vskip10pt

Previously, we have studied\cite{GT2} how the ratio between the transverse $\Gamma_{\perp}$ and longitudinal
$\Gamma_{\parallel}$ components of the plasmon resonance half-width depends on the degree of ellipsoid's
oblateness or prolateness for frequency region that is located higher or lower than the characteristic
frequency of an electron reflections between particle walls.

Up to now, we considered only nonradiative processes, when the electron scattering in the MN
dissipates oscillation energy into heat. Below we dwell shortly on the radiative processes.

\section{Radiative damping}

The problem of a damping of the electron energy due to the radiation
of a portion of the collective electron oscillation energy into the optical
far field has been extensively studied in the literature.\cite{KV,SFW,MW,DSP,LPG}

As we have seen from the above sections, the plasmon linewidth for small MNs ($R\leq 100\AA$)
essentially depends on electron collisions with a particle surfaces (dissipative damping).
However, for large MNs ($R\gg 100\AA$), our mechanism do not account for the dissipation of
the electron energy due to inverse transformation of localized plasmons into propagating
electromagnetic radiation (radiative damping).

The relative contributions from radiative damping through the resonant scattering and absorption are also strongly
depend on the particle size. In particular, it is known\cite{BH,KV} that the plasmon absorption is the only process
in small particles, whereas both the absorption and the scattering are present in large particles, with the latter
becoming more dominant as the particle size increases. The phenomenon is based on interplay of the usual dissipative and
radiative damping, where the latter is related to inverse transformation of localized resonant plasmons into scattered light.

The charges, under approach of classical electrodynamics, radiate when they move with acceleration.
To calculate the line broadening that is entirely caused by an increase of $\Gamma_{rad}$ due
to the radiant effect, we will use the time dependence of a classical dipole oscillator.
The force of a decelerative radiation of a dipole under an inner electric field (see Eq.~(\ref{eq 3}))
can be presented as
\be
 \label{eq 49}
  {\bf F}_{\rm rad}(t) = -\frac{2e}{3c^3}\sqrt{\epsilon_m}\frac{{\bf \dddot{d}}(t)}{[1+L(\epsilon/\epsilon_m-1)]^{-1}}.
   \ee
In the case of a medium with $\varepsilon_m=\varepsilon=1$, Eq.~(\ref{eq 49})
transforms into those well-known from the classical electrodynamics.\cite{J}
The linewidth due to the radiative damping of dipole vibrations is connected to the ${\bf F}$ by means of
\be
 \label{eq 50}
  \Gamma_{\rm rad} = \frac{e}{m}{\rm Im} \left[\frac{{\bf F}_{\rm rad}(t)}{{\bf \dot{d}}(t)}\right]N,
   \ee
where $N$ is the number of free electrons in the MN.
Supposing ${\bf d}(t)={\bf d}_0\exp{(-i\omega t)}$,
we obtain for $j$-th component of a radiative linewidth the following expression:
\be
 \label{eq 51}
  \Gamma_{j,\rm rad} = \frac{2}{3}\frac{e^2\omega^2}{mc^3}N\,{\rm Im}
   \left[\frac{\sqrt{\epsilon_m}}{[1+L_j(\epsilon_{jj}/\epsilon_m-1)]^{-1}}\right],
    \ee
or with accounting for the dielectric matrix properties given by Eq.~(\ref{eq 7}),
 Eq.~(\ref{eq 51}) is reduced to the form
\be
 \label{eq 53}
  \Gamma_{j,\rm rad}(\omega) = \frac{2}{3}\frac{e^2 \omega^2}{mc^3}N
   \frac{L_j\epsilon''_{jj}(\omega)}{\epsilon_m^{1/2}}.
    \ee
Taking into account the expression (\ref{eq 6}) for $\epsilon''$, we get
\be
 \label{eq 54}
  \Gamma_{j,\rm rad}(\omega) = \frac{8\pi}{3}\frac{e^2 \omega}{m c^3}N
   \frac{L_j\sigma'_{jj}(\omega)}{\epsilon_m^{1/2}}.
    \ee

Different relations can be employed for $\sigma'_{jj}$, depending on the frequency regime.
For instance, in the highfrequency limit, when Eq.~(\ref{eq 31}) can be applied, we obtain for two
component of the linewidth  of a spheroidal MN embedded in a medium with $\epsilon_m$ the following equation
\be
 \label{eq 55}
  \Gamma_{{\|\choose\bot},{\rm rad}}=\frac{3}{4}\frac{e^2}{m c^3}
   \frac{\omega^2_{\rm pl}}{\omega}N\frac{\upsilon_F}{R_{\bot}}  
    \frac{L_{\|\choose\bot}}{\epsilon_m^{1/2}} {\eta(e_p)\choose\rho(e_p)},
     \ee
where the functions $\eta(e_p)$ and $\rho(e_p)$ are given by Eqs.~(\ref{eq 32}) and ~(\ref{eq 33}), respectively.
In the case of a spherical particle, Eq.~(\ref{eq 55}) with accounting for Eq.~(\ref{eq 11}), can be rewritten as
\be
 \label{eq 56}
  \Gamma_{\rm rad, sp}=\frac{1}{6}\frac{e^2 \omega_{\rm pl}}{m c^3}N\frac{\upsilon_F}{R}
   \frac{(1+2\epsilon_m)^{1/2}}{ \epsilon_m^{1/2}}.
    \ee

The increase in linewidth from radiative damping is proportional to the surface area of the MN. 
The effect is weaker in a higher dielectric constant of environment.

The estimations of the $\Gamma$ for the spherical Au, Ag, Cu and Na particles with $2R=200\AA$
embedded in water ($\epsilon_m=1.78$) give: 0.942, 0.927, 1.82, and 1.13~meV, respectively.
It is known from Mie scattering theory\cite{KV} that only 1.5$\%$ of the total
damping rate in $2R = 200~\AA$ metallic spheres is due to the radiative decay.

In order to take into account radiative damping together with collisions of free carriers with the MN
surface, the effective collision frequency $\Gamma_{\rm eff}=\Gamma+\Gamma_{\rm rad}$ must be introduced.

\section{Summary}
	
We use the kinetic approach to study the plasmon resonance linewidth for the metal nonspherical nanoparticles
embedded in any dielectric media. It enables one to calculate the linewidth in the case that the free electron
path is much larger than the particle size and the scattering from the particle surfaces plays a dominate role.

The general formula is proposed for a damping rate or a decay time due to electron scattering from the
bulk and particle surfaces. By means of this formula one will have a possibility to evaluate
the linewidth directly through the tensor of polarizability of the MN.

The electron surface-scattering contribution to the plasmon damping in a simple case of a spherical metal nanoparticles
is studied in detail. It is clearly shown that the resonance plasmon linewidth oscillates as the
particle radius increases. The oscillating terms represent an important correction to the linewidth, especially at a small
particle radii. This result for the Na nanoparicles is in well agreement with the numerical time-dependent
local density approximation calculations.

With changing the MNs shape from spherical to the spheroidal one, the single plasmon resonance splits into
two components: the longitudinal and a transverse one to the spheroid rotation axis. Both components of
the linewidth oscillate with the axial ratio altering. The period of these oscillations depends on the particle
volume and is enhanced with an increase in the axial ratio. The amplitude of oscillations enhances also with the
contraction of the particle volume. The behavior of the linewidth extrema  are studied for both components of an
oblate and a prolate MNs with different volumes.

The resonance plasmon damping in the {\em oblate} Na nanoparticle
was found stronger along the spheroid revolution axis than the one across this axis.
For the {\em prolate} Na nanoparticle, on the contrary, the damping along the revolution axis
was weaker than the one across this axis.

The size-dependent oscillations of the linewidth also depend on the dielectric constant.
For the first time, we detect that the oscillations of the surface plasmon resonance may
occur also with an increasing of the dielectric constant of the surrounding medium.
The oscillations are well pronounced for nanoparticles with the
small radii and disappeared for NP with a larger radii.
The magnitude of these oscillations is the greater the smaller particle is and enhances markedly with $\epsilon_m$.

The effects of both the particle shape and the environment on the plasmon resonance linewidth are illustrated
by the example of the Na nanoparticles with a different radii.

The contribution of the radiative plasmon decay is discussed as well.

Our theoretical results should be important for the analysis of the transport
and optical properties of MNs under an exposure of short and strong laser excitations.

\end{document}